\documentclass{article}
\usepackage{mrs2005,epsfig}
\def\lesssim{\mathrel{\hbox{\rlap{\hbox{\lower4pt\hbox{$\sim$}}}\hbox{$<$}}}}
\def\gtrsim{\mathrel{\hbox{\rlap{\hbox{\lower4pt\hbox{$\sim$}}}\hbox{$>$}}}}
\def\sun{\hbox{$\odot$}}
\begin{document} 
\title{EVOLUTION OF LYMAN BREAK GALAXIES FROM Z=5 TO 3}

\author{Ikuru Iwata}
\affil{Okayama Astrophysical Observatory, National Astronomical Observatory of Japan} 
\author{Kouji Ohta, Masataka Ando, Gaku Kiuchi}
\affil{Department of Astronomy, Faculty of Science, Kyoto University, Japan}
\author{Naoyuki Tamura}
\affil{Department of Physics, University of Durham, UK}
\author{Masayuki Akiyama and Kentaro Aoki}
\affil{Subaru Telescope, National Astronomical Observatory of Japan}

\begin{abstract} 
We report the updated UV luminosity function (LF) of Lyman break galaxies 
at $z \sim 5$. Combining with lower redshift data, we found that the 
evolution of UV LF is differential depending on the UV luminosity. 
With results from clustering analysis, optical spectroscopy and SED fitting, 
it is suggested that this differential evolution might be 
understood as a consequence of biased galaxy evolution. 
\end{abstract} 
 
\section{Introduction} 

In the last ten years many deep surveys using $HST$ and 
ground-based facilities have been carried out, and 
we have gradually enlarged our knowledge on the nature of 
galaxies in the early universe. 
The Lyman break technique, which was first developed by 
C. Steidel and his co-workers, is one of the simplest and 
most successful methods to extract galaxies at target redshift 
ranges from deep survey images. 
Past studies have successfully constructed a large samples of 
Lyman break galaxies (LBGs) mainly at $z \lesssim 3$, and 
follow-up studies have gradually revealed their properties 
(e.g., Steidel et al. 2003; Sawicki and Yee 1998; Papovich et al. 2001; 
Shapley et al. 2003; Adelberger et al. 2005).

A unique combination of the large mirror, wide-field camera and excellent 
image quality of Subaru telescope enabled us to proceed toward 
higher redshift, approaching to the epoch of galaxy formation. 
In 2001 our team pointed the telescope to the HDF-N/GOODS-N area 
and made multi-band optical imaging aiming at 
the survey of LBGs at $z \sim 5$. 
We have successfully detected more than 300 LBG candidates 
with $I_c$ $\leq 26.0$ \footnote{All magnitudes are in AB} 
as the world's first large sample 
of LBGs at $z \sim 5$ (Iwata et al. 2003). 
Follow-up spectroscopy using FOCAS revealed that relatively 
bright 7 objects ($I_c \leq 25.0$) are 
indeed at $z \sim 5$ (Ando et al. 2004). 
Other teams using Subaru have also made systematic census of 
LBGs at $z \sim$4--6 (e.g., Ouchi et al. 2004; Shimasaku et al. 2005). 
Although the surveyed area is quite narrow, the Hubble 
Ultra Deep Field project put constraints on the number 
density of LBGs at $z\gtrsim 6$ (e.g., Bouwens et al. 2004, 2005; 
Bunker et al. 2004).

In this contribution, we would like to report the 
updated UV luminosity function for LBGs at $z \sim 5$, 
as well as some results from follow-up observations, 
clustering analysis and SED fitting using public released 
data from GOODS (Giavalisco et al. 2004). 
A discussion on the evolution of LBGs is also presented.

\section{Sample} 

Optical imaging data from which we have 
selected sample of LBGs at $z \sim 5$ were 
taken with Subaru / Suprime-Cam (Miyazaki et al. 2002), 
which has a $34' \times 27'$ field of view. 
There are two target fields, namely the Hubble Deep 
Field - North (HDF-N) and J0053+1234 (Cohen et al. 1996). 
The HDF-N data were taken in 2001 by us and the UH group 
(Capak et al. 2004; taken from archive system SMOKA). 
J0053+1234 data have been taken from 2002--2004 by us. 
The limiting magnitudes of images ($1.2'' \phi$ aperture, 
5$\sigma$) for the HDF-N 
and J0053+1234 were 28.2($V$), 26.9($I_c$), 26.6 ($z'$) 
and 27.8($V$), 26.4($I_c$), 26.2($z'$), respectively.

The color selection criteria for 
LBGs at $z \sim 5$ we have adopted were 
same as those described in Iwata et al. (2003); 
$V-I_c\geq1.55$ and $V-I_c\geq7(I_c-z')+0.15$.
The number of LBG candidates are 617 ($z'<26.5$) 
for HDF-N and 246 ($z'<25.5$) for J0053.

\section{UV Luminosity Function and Optical Spectroscopy}

In figure 1 we show the UV luminosity function for 
our $z\sim5$ LBG samples, 
as well as those from Keck Deep Fields (Sawicki and 
Thompson 2005) for $z \sim 3$--4 and UDF-Parallel 
(Bouwens et al. 2004) for $z \sim 6$. 
From redshift 5 to 3, the number density of UV 
luminous objects does not show significant change. 
However, the number density of faint $L < L^\ast$ galaxies 
is increasing by a factor of $\approx$5, although 
our $z \sim 5$ sample is more than 1 mag shallower than 
those for lower redshifts. 
This indicates that the number density evolution is 
differential depending on the UV luminosity 
(see also a contribution by M. Sawicki in this 
volume).

On the other hand, UV LF for LBGs at $z \sim 6$ might show 
significant drop in the bright part (e.g., Bouwens et al. 2004; 
Bunker et al. 2004). If it is the case, 
drastic evolution may have occurred from $z \sim 6$ to 5. 
Note that $z \sim 6$ is considered to be an epoch of 
the end of the cosmic reionization (Fan et al. 2001; Cen 2003).

%  
% Figure 1 
% 
\begin{figure}  
\vspace*{1.25cm}  
\begin{center}
\epsfig{figure=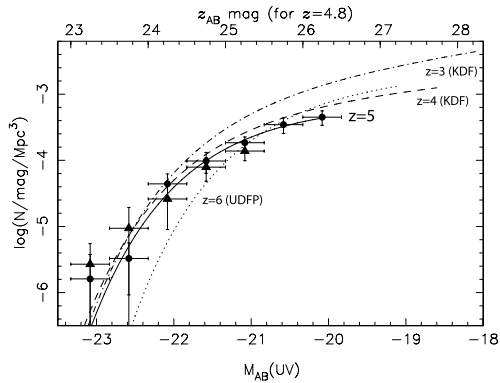,width=6.5cm}  
\end{center}
\vspace*{0.25cm}  
\caption{UV luminosity function (LF) of LBG samples 
from $z \sim 6$ to $z \sim 3$. UV LFs for $z\sim5$ 
are for HDF-N(circles) and J0053+1234(triangles). 
Schechter function fits for other redshifts are from 
$z \sim 6$: Bouwens et al. (2004), $z \sim 3$ and 4: 
Sawicki and Thompson (2005).
} 
\end{figure} 

%\section{Optical Spectroscopy}

We have made optical spectroscopy for limited number of 
sample LBGs with Subaru/FOCAS (Kashikawa et al. 2002). 
So far nine luminous ($L>L^\ast$) LBGs have been identified 
as objects at $z \sim 5$ (Ando et al. 2004; Ando et al. in prep.). 
There is significant absence of strong Ly$\alpha$ emission 
for luminous LBGs at $z \sim 5$, which is not apparent 
for $z \sim 3$ LBGs (e.g, Shapley et al. 2003). 
In the spectra of those LBGs, we also see strong interstellar 
low-ionization metal lines such as SiII, SiIV and CII. 
These features might be attributed to (1) relatively 
dusty and metal-rich environment of luminous LBGs, or 
(2) massive neutral gas reservoirs around the star-forming 
galaxies. 
More detailed descriptions of spectroscopic results are 
presented in a separate contribution by M. Ando.

\section{Clustering and Stellar Populations}

We derived the two-point correlation function for our 
$z \sim 5$ LBGs. 
Clustering signals were detected (figure 2, 
left), 
and there seems to be a trend that UV luminous 
objects showed stronger clustering than fainter ones, 
as it has been reported for $z \lesssim 3$ LBGs (e.g., 
Giavalisco and Dickinson 2001; Adelberger et al. 2005). 
Correlation lengths of $z \sim 5$ LBGs (figure 2, right) are 
slightly larger than LBGs at lower redshifts and are 
comparable to 
Distant Red Galaxies and Sub-mm galaxies, many of both populations 
are considered to be active star forming galaxies (Iwata et al. 
2005; Blain et al. 2004). It is suggested that UV luminous LBGs 
at $z \sim 5$ are hosted by massive dark matter halos, 
which could grow into host halos of giant galaxies 
in the present universe.

%  
% Figure 2
% 
\begin{figure}  
\vspace*{1.25cm}  
\begin{center}
%\plottwo{./figures/pg_angcluster02.eps}{./figures/pg_r0_01.eps}  
\plottwo{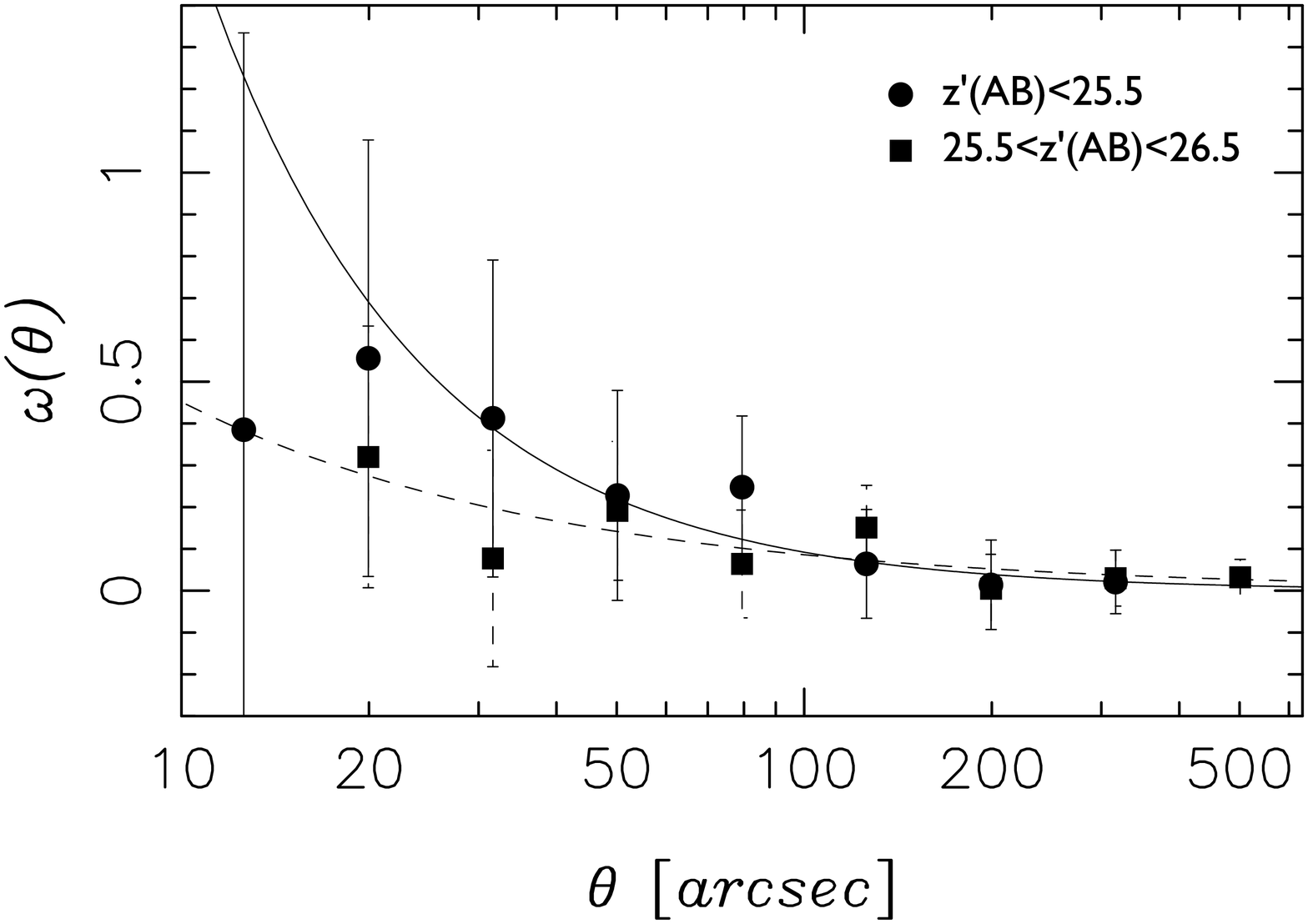}{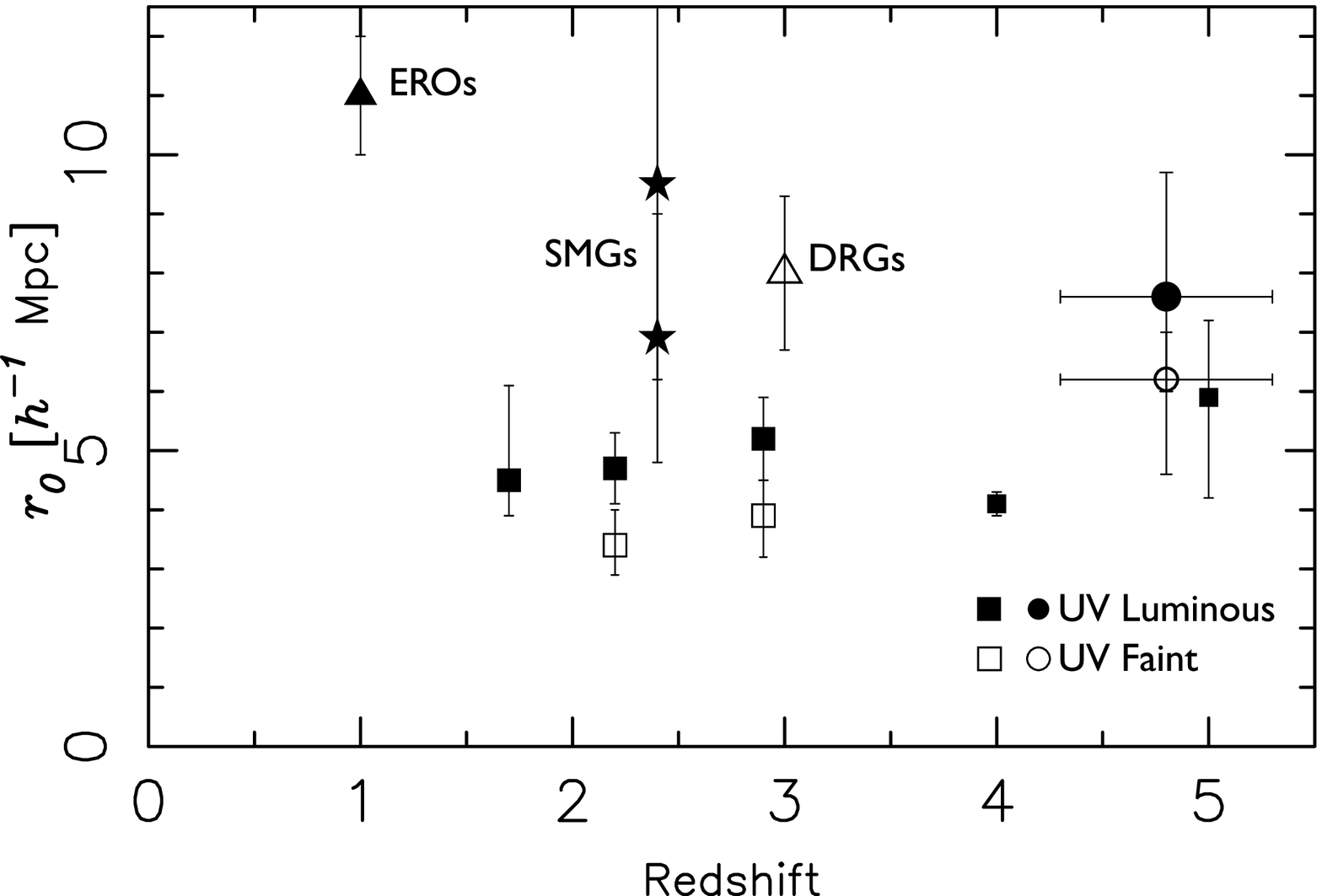}  
\end{center}
\vspace*{0.25cm}  
\caption{(Left)Correlation functions of luminous (circle) and 
faint (square) LBGs at $z \sim  5$. ~
(Right): Comparison of correlation lengths for various 
populations at different $1 \lesssim z \lesssim 5$. 
Data points other than ours at $z=4.8$ are from 
Adelberger et al. (2005; LBGs at $z \sim $ 2--3), 
Ouchi et al. (2004; LBGs at $z \sim$4--5), 
Miyazaki et al. (2004; EROs), Blain et al. (2004; SMGs)
and Daddi et al. (2003; DRGs).
} 
\end{figure} 

%\section{Stellar Populations} 

The cross-identification of our sample LBGs with the Spitzer/IRAC images 
in the public data release (DR1) from the GOODS have been made. 
Although the survey depths and coverage are still limited, about 100 objects 
were detected in IRAC channel 1 and/or 2. 
In figure 3 (left) we show a correlation between IRAC flux and rest-frame 
UV-to-optical colors. The absence of luminous LBGs 
with blue colors is clearly seen. 
This trend might be partly due to the dusty environment in luminous 
LBGs. However, it might be a natural consequence of the existence of 
evolved stellar populations, since IRAC channel 1 and 2 trace 
rest-frame $\approx$6,000--8,000\AA. 

We have also tried to obtain 
$K'$-band images for limited number of sample galaxies with Subaru/CISCO.
With these near- and mid-IR data we have carried out SED fitting. 
For calculating flux densities of sample galaxies in IRAC 
we tested both $1.''6$ aperture (same as used in optical data) 
photometry with corrections for larger PSFs of IRAC data 
and \verb+mag_auto+ in SExtractor (Bertin and Arnouts 1996), and 
found that the differences in estimates are well within 68\% 
confidence intervals. 
Model spectra were generated by version 2 of PEGASE (Fioc \& 
Rocca-Volmerange 1997). Two simplified star formation histories 
were calculated, namely, constant star formation rate (SFR) and 
exponentially-decaying star formation ($\tau=$10--1,000 Myr). 
One of the results from this preliminary analyses 
with constant SFR models is shown in figure 3 (left). 
Fairly tight correlation between the IRAC (rest-frame optical) flux densities 
and stellar mass estimates is found, and there are some optically luminous 
galaxies which have 
already accumulated $\gtrsim 10^{10} M_{\sun}$ stellar mass at $z\sim 5$. 
We also found that there seems to be a tendency that luminous 
LBGs are subject to slightly larger dust attenuation, 
which might be qualitatively consistent with results from spectroscopy, 
but the degeneracy of dust attenuation and stellar ages in 
the current data set prevents us to reach the definitive conclusion 
at the moment.

%  
% Figure 3
% 
\begin{figure}  
\vspace*{1.25cm}  
\begin{center}
%\plottwo{./figures/colormag_iraf.eps} {./figures/pg_sedfit_summary01r_const_ap_sel.eps} 
\plottwo{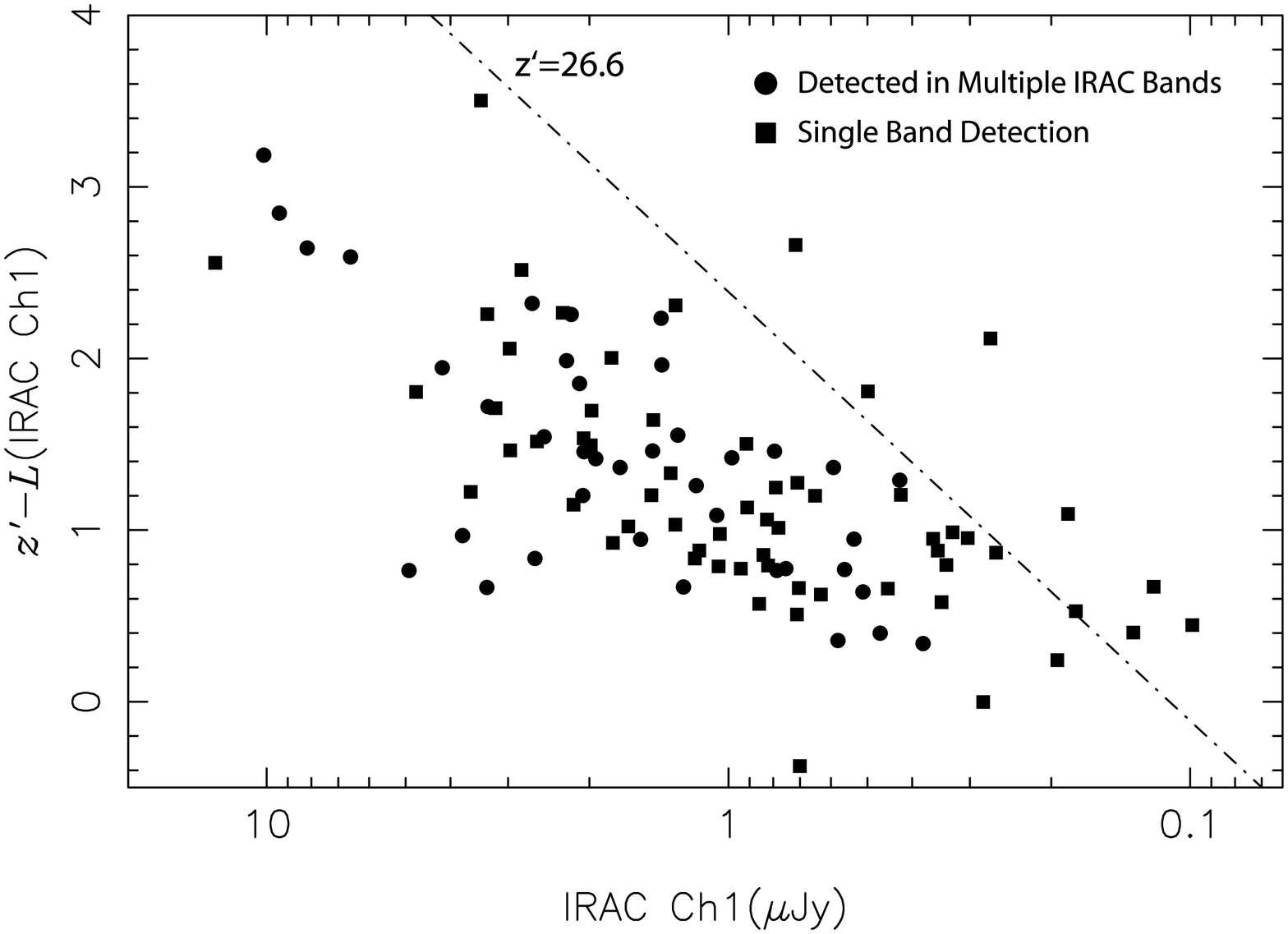} {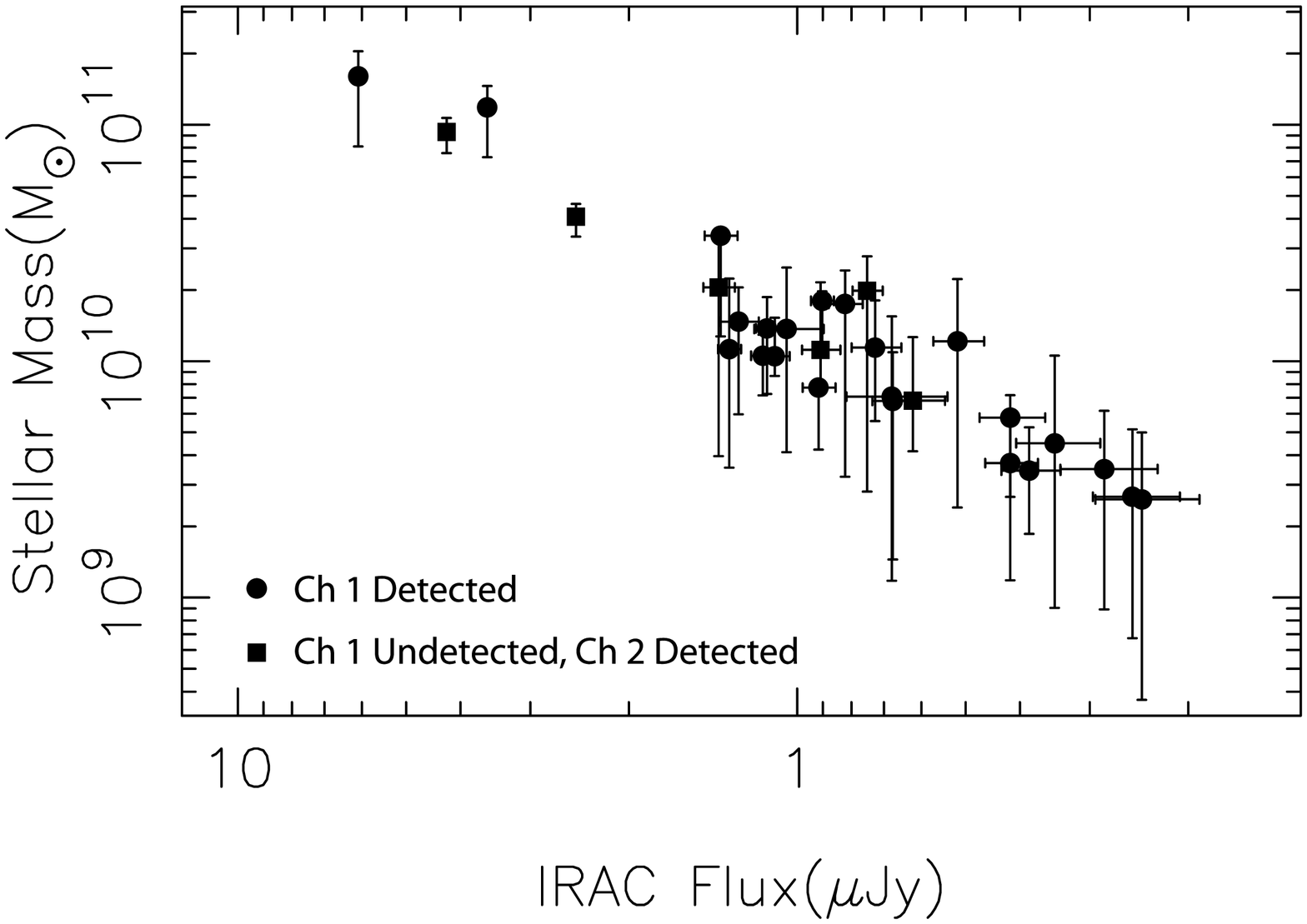} 
\end{center}
\vspace*{0.25cm}  
\caption{
(Left): IRAC flux densities (in channel 1 or 2) and $z'-L$ colors, which corresponds 
to the rest-frame UV-to-optical colors.
(Right): one example of results of SED fitting for IRAC-detected LBGs at $z \sim 5$. 
stellar mass estimates are plotted along with IRAC flux densities.
} 
\end{figure} 

\section{A Possible Scenario: Biased Galaxy Evolution}

These observational results coming from deep imaging and 
spectroscopy as well as multi-wavelength data might open out 
a new perspective of the galaxy evolution at the universe age less 
than 2 Gyr. 
If we assume that actively star-forming galaxies (=UV luminous 
galaxies) have larger amount of gas (suggested from optical spectroscopy) 
and are hosted by more massive DM halos (suggested from the difference 
in clustering amplitude), the differential evolution of UV LF can be 
closely connected to the biased galaxy evolution scenario: 
Star formation started selectively in rare peaks of mass density at 
$z > 5$. Such most massive DM halos host large amount of baryons as well, 
and the active star formation would be ignited (resulting fairly 
large amount of dust and metals). As time passes the clustering of 
DM halo grows and star formations gradually take place within less 
massive DM halos, which are more common in the universe. This would 
result in the increase of fainter objects by $z \sim 3$. 

Moreover, this scenario could be put into a larger view of the 
galaxy evolution throughout the cosmic time. 
From recent studies it seems that the star formation density 
in the universe gradually increase from $z \sim 6$, reached 
to the highest level around $z=3$--2 then slowly declines until the 
current epoch(e.g., Bunker et al. 2004). 
The increase at higher redshift should be 
caused by the differential UV LF evolution, and the decline at 
low-z would be tightly connected to the ``down-sizing'' effect 
(Cowie et al. 1996; Kodama et al. 2004); i.e., the massive galaxies 
stops star formation in earlier epochs and less massive ones continue 
to form stars. With more thorough investigation for high-z 
biased galaxy evolution (as we describe below), we would be able to 
unify it with the ``down-sizing'' effect and depict an overall evolution 
viewgraph of galaxies from the cosmic age of 1 Gyr to 13 Gyr, as 
schematically illustrated in figure 4.

%  
% Figure 4
% 
\begin{figure}  
\vspace*{1.25cm}  
\begin{center}
\epsfig{figure=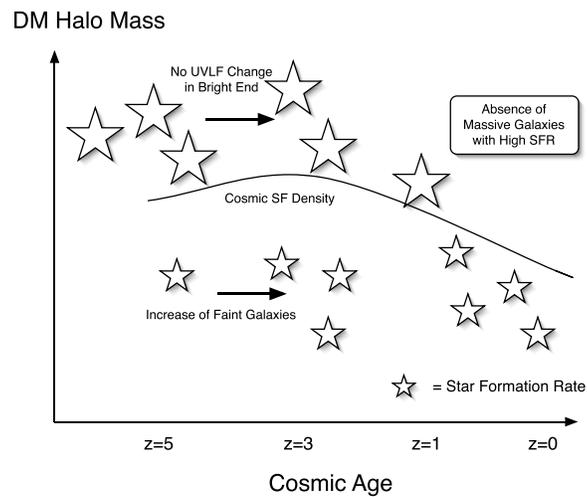,width=8cm}  
\end{center}
\vspace*{0.25cm}  
\caption{A schematic illustration of the unified view of the 
``biased galaxy evolution'' at $z \gtrsim 3$ and ``down-sizing'' 
of $z \lesssim 2$.} 
\end{figure} 

Although this scenario qualitatively explains all of our findings 
consistently, that does not mean it is the unique solution. 
There are some alternative possibilities: there might be a change 
in the properties of starburst (frequency, intensity, or 
duty cycle of burst-quiescent loops) from $z\sim5$ 
to $z\sim3$, or differential dust properties might 
be also able to reproduce some of observed trends (Sawicki and 
Thompson 2005). 
%For example, if the frequency of starburst cycle becomes sparse 
%at $z \sim 3$, the fraction of galaxies with low UV luminosity 
%increases and the faint-end slope of UV LF becomes steeper.
%The difference in dust properties might be also able 
%to explain the differential evolution. 
We should also consider that LBGs are by definition 
UV luminous star-forming galaxies with relatively small amount 
of dust attenuation, and some part of 
LBGs at higher redshift would be missed when they are observed  
at lower redshift, due to passive evolution or larger amount 
of dust (like sub-mm galaxies).
Although we think that a new perspective for a comprehensive 
understanding of galaxy evolution throughout the cosmic time 
began to open out for us, 
various kinds of improved data would be required to go 
further.
Deeper images in optical, infrared and sub-mm 
wavelengths for the construction of the complete samples and 
more reliable stellar population / dust amount estimations 
through SED fitting, and much more spectroscopic identifications 
to obtain insights on matallicities and gas kinematics within galaxies. 
We expect that such data would be 
helpful for the improvement of 
ingredients of cosmological numerical simulations, 
to consistently predict properties of star-forming galaxies 
at wide range of redshifts.

%\section{Conclusions} 
 
\acknowledgements{
We thank M. Sawicki for providing their results for UV LF at $z \sim$2--4 
and valuable discussions. 
A part of this study is based on data collected at Subaru Telescope and obtained 
from the SMOKA science archive at Astronomical Data Analysis Center, 
which are operated by the National Astronomical Observatory of Japan.
}

\vfill 
\end{document}